\documentclass[a4paper,10pt,twocolumn,aps,prl,nolongbibliography,superscriptaddress,showpacs,showkeys,amsmath,amssymb,floatfix,nofootinbib]{revtex4-1}
\usepackage{lmodern}
\pdfoutput=1

\usepackage[T1]{fontenc}
\usepackage[utf8]{inputenc}
\usepackage[usenames,dvipsnames]{xcolor}
\usepackage{graphicx}
\usepackage[dvipsnames]{xcolor}
\usepackage[unicode=true,pdfusetitle,
 bookmarks=true,bookmarksnumbered=true,bookmarksopen=true,bookmarksopenlevel=1,
 breaklinks=false,pdfborder={0 0 0},backref=false,colorlinks=true]
 {hyperref}
\hypersetup{
 citecolor=blue,filecolor=blue,linkcolor=blue,urlcolor=blue}
\makeatletter

\renewcommand{\[}{\begin{equation}}
\renewcommand{\]}{\end{equation}}

\usepackage{array}
\makeatother

\begin{document}

\title{Direct  detection of gravitational waves can measure \\ the time variation of the Planck mass}

\author{Luca Amendola}
\affiliation{Institut für Theoretische Physik, Ruprecht-Karls-Universität Heidelberg,
Philosophenweg 16, 69120 Heidelberg, Germany}

\author{Ignacy Sawicki}
\affiliation{CEICO, Institute of Physics of the Czech Academy of Sciences, Na Slovance 2, 182 21 Praha 8, Czechia}

\author{Martin Kunz}
\affiliation{Départment de Physique Théorique and Center for Astroparticle Physics,
Université de Genève, Quai E. Ansermet 24, CH-1211 Genève 4, Switzerland}

\author{Ippocratis D.~Saltas}
\affiliation{CEICO, Institute of Physics of the Czech Academy of Sciences, Na Slovance 2, 182 21 Praha 8, Czechia}

\begin{abstract}

 The recent discovery of a $\gamma$-ray counterpart to a gravitational wave event has put extremely stringent constraints on the  speed of gravitational waves at the present epoch. In turn, these constraints place strong theoretical pressure on potential modifications of gravity, essentially allowing only a conformally-coupled scalar to be active in the present Universe. In this paper, we show that direct detection of gravitational waves from optically identified sources can also measure or constrain the strength of the conformal coupling in scalar--tensor models through the time variation of the Planck mass.

As a first rough estimate, we find that the LISA satellite can measure the dimensionless time variation of the Planck mass (the so-called parameter $\alpha_M$) at redshift around 1.5 with an error of about 0.03 to 0.13, depending on the assumptions concerning future observations. Stronger constraints can be achieved once reliable distance indicators at $z>2$ are developed, or with GW detectors that extend the capabilities of LISA, like the proposed Big Bang Observer. We emphasize that, just like the constraints on the gravitational speed, the bound on $\alpha_M$ is independent of the cosmological model.

\end{abstract}

\maketitle

\section{Introduction}

The recent almost simultaneous observation of gravitational waves (GWs) and $\gamma$-rays from neutron star binaries \cite{TheLIGOScientific:2017qsa} has opened a new era of multi-messenger astronomy. One of the most interesting aspects of this event has been the first-ever precise measurement of the GW speed $c_T$, confirming what General Relativity (GR) predicted, namely that the present value of $c_T$ equals the speed of light $c$ to an astonishing precision, $|c_T/c-1|\lesssim 10^{-15}$ \cite{2041-8205-848-2-L13}. 

Modified gravity theories that couple extra scalar or vector degrees of freedom to curvature, change the propagation of gravitational waves in one of two ways: they allow in general for a propagation speed $c_T$ different from $c$, and make the effective Planck mass $M_*$ time and/or position dependent. 
At the same time, the metric sourced by massive bodies is modified by the so-called gravitational slip, see e.g for earlier and more recent discussions \cite{PhysRevD.77.103513,Saltas:2014dha,Sawicki:2016klv,Amendola:2012ys}. In this context, we use as a working definition of modified gravity the one we chose in the above works, that is, any theory which modifies the propagation of GWs. 

For example, scalar-tensor theories with second-order equations of motion find their most general formulation in the so-called Horndeski Lagrangian \cite{Horndeski:1974wa,Deffayet:2011gz,Kobayashi:2011nu}. The four free functions that enter this Lagrangian, often denoted as $G_2,G_3,G_4,G_5$, are functions of the scalar field, and are associated with two sharply separate sectors: the first one ($G_2,G_3$) only affects the scalar-field evolution, while the second in principle couples non-minimally (i.e.\ beyond the standard gravitational coupling) the scalar field to gravity ($G_4,G_5$), the latter affecting both the scalar's evolution and GWs \cite{DeFelice:2011bh} \footnote{Whenever $\partial {G_{4}(\phi, X)}/{\partial X}$ or $G_5\neq0$, the speed of GWs is affected, otherwise, this sector reduces to the standard, conformal coupling to curvature. We should also add here that, the so-called Kinetic Gravity Braiding term described by the free function $G_{3}(\phi,X)$ does not modify the propagation of GWs not does it produce gravitational slip and hence does not fall into our chosen definition for modified gravity. Nonetheless, it does still modify non-trivially the constraint structure of gravity compared to GR owing to the kinetic mixing between the scalar and gravity and thus changes the effective Newton's constant for perturbations.}. A similar structure exists for general vector-tensor theories: Einstein-Aether \cite{Jacobson:2000xp} and generalized Proca \cite{Heisenberg:2014kea,Tasinato:2014eka}.

The complete effect of gravity modification at the level of linear perturbations can be described by a small set of functions of time alone \cite{Gubitosi:2012hu,Gleyzes:2013ooa,Bloomfield:2013efa}. For example, Ref.~\cite{Bellini:2014fua} makes the choice, which we will adopt in this paper, to parametrize the general Horndeski theory with four functions $\alpha_M,\alpha_T,\alpha_B,\alpha_K$, which in turn depend on the $G_i$ functions appearing in the Lagrangian. From those, only the running of the Planck mass $\alpha_M$  and the excess in the tensors' speed $\alpha_T$,
\begin{align}
& \alpha_M= \mathcal{H}^{-1} \frac{d\ln M_*^2}{dt}, \\
& \alpha_T \equiv c^2_T - 1,
\end{align}
express the non-minimal interaction with gravity. In turn, the non-minimal interaction between the scalar and curvature can be separated into a conformal part (i.e., the sector that can be absorbed into a conformal rescaling of the metric which redefines the Planck mass), and a non-conformal part (the sector which changes the speed of GWs). 
An important consequence is that the scalar field is only minimally coupled to gravity if, and only if, $\alpha_M=\alpha_T=0$.  In this case, gravity is no longer modified and GWs propagate as in standard GR. A similar parametrization can be adopted for other theories, such as beyond Horndeski \cite{Zumalacarregui:2013pma,Gleyzes:2014dya} and vector-tensor models, with the same functions $\alpha_M, \alpha_T$ describing fully the non-minimal interaction with gravity \cite{Lagos:2017hdr}.  

In view of the above parametrisation, the LIGO event \cite{TheLIGOScientific:2017qsa}, then, tells us that $\alpha_T=0$ with great precision. This implies that any non-conformal coupling between the scalar and curvature vanishes at the present epoch \cite{Creminelli:2017sry,Ezquiaga:2017ekz,Sakstein:2017xjx,Baker:2017hug}. (see also \cite{Bettoni:2016mij,Lombriser:2016yzn}). While this constraint is enough to forbid any sort of non-minimal coupling in vector-tensor theories \cite{Baker:2017hug,Amendola:2017orw}, for scalar-tensor theories a conformal non-minimal coupling is still allowed. It is also possible to use this event to constrain $\alpha_M$, but the bound is extremely weak \cite{Arai:2017hxj}.
	
In this paper, we show that GWs from sources with identifiable redshift can also measure and constrain the evolution of the effective Planck mass, i.e.\ constrain the second modified-gravity parameter, $\alpha_M$ strongly, and with this, the remaining conformal coupling for scalar-tensor theories. In this way, GWs can constrain or rule out the entire modified-gravity sector of both vector-tensor and scalar-tensor models.

One generically expects that if there is such a conformal coupling of gravity, the model must feature screening so that precision tests of gravity do not already rule it out. This screening mechanism would act as to suppress the Solar-System value of $\alpha_M$, which is essentially the rate of change per Hubble time of the gravitational constant, compared to that in the wider cosmology. The present and local value of $|\alpha_M|$ can indeed be constrained to be less than 0.01$\div$0.03 in the laboratory and in the Solar System (see for instance a recent summary of results and a  positive detection in \cite{2017arXiv170701866B}). A  cosmological  constraint from Big-Bang nucleosynthesis (BBN) is also a stringent one, $|G_\text{BBN}/G_0-1|\lesssim 0.2$ \cite{2004PhRvL..92q1301C}.  As will be shown in the following, the completely independent test we propose here can reach similar or even better sensitivity.

The idea of using GWs to test $\alpha_M$ and $\alpha_T$ was put forward for the first time in \cite{Pettorino:2014bka}, where it was shown that $B$-modes created by primordial GWs in the polarized Cosmic Microwave Background (CMB) sky can in principle constrain both quantities. The Planck's CMB analysis \cite{Ade:2015rim} produced, for some  classes of functional parametrization of $\alpha_M(t)$,  errors around 0.05 at 95\% confidence level for the present value of $\alpha_M$. These errors, however, depend on the assumption of a standard cosmological model and, in particular, of a $\Lambda$CDM background. Therefore, these are tests of structure formation for particular modified gravity models, rather than direct tests of generic modifications of gravity. 

In contrast, we shall emphasise that the method we propose here is independent of the underlying cosmological model and of the precise model of modified gravity. Another advantage with respect to CMB or BBN constraints is that one can in principle map the evolution of $\alpha_M$ in an extended redshift range from today to $z\approx 8$.

\section{GW propagation}
We consider a flat Friedmann-Robertson-Walker (FRW) spacetime with scale factor $a$ and conformal Hubble function $\mathcal{H}$.
As it has been shown in \cite{Saltas:2014dha}, in such a cosmological background the GW amplitude $h$ in any modified gravity theory which does not give gravitons a mass, obeys the equation
\[
\ddot{h}+(2+\alpha_{M})\mathcal{H}\dot{h}+c_T^2 k^{2}h=0,\label{gweq}
\]
where the dot stands for a derivative with respect to conformal time $t$, $c_T$ is the speed of GWs. The quantity $\alpha_M$, already defined in the Introduction, expresses the time variation of the time-dependent effective Planck mass $M_*$ (see \cite{Bellini:2014fua}). $M_*^2$ is defined as the normalization of the kinetic term for the metric fluctuations $h$ in the action for perturbations. For example, in the simple case of a Brans-Dicke gravity with parameter $\omega$, one finds $\alpha_M=1/(1+\omega)$. In more general models of modified gravity, the GW equation is also of this form, see Refs~\cite{Saltas:2014dha,Nishizawa:2017nef} for details.

The GW event reported in Ref. \cite{TheLIGOScientific:2017qsa} has shown that $c_T=1$ with extreme precision, at least for the present Universe. Here we would like to investigate the observable effects of $\alpha_M$ on the GW signal, remembering that, fixing $\alpha_M,\alpha_T$, as already mentioned, amounts to completely fixing the non-minimal scalar-tensor interaction.  

Let us define the field $v\equiv M_{*}ah$. This quantity  obeys the equation of motion
\[
\ddot{v} + k^2 v - \mu ^2 v=0, \label{eq:sho}
\]
with tachyonic mass $\mu$ of order $\mathcal{H}$, and given by $4\mu^2 \equiv (2+\alpha_M)^2\mathcal{H}^2 + 2(2+\alpha_M)\dot{\mathcal{H}} + 2\dot\alpha_M \mathcal{H}$. So, provided that the wavelength of the GW is subhorizon, $k\gg \mathcal{H}$, $v$ evolves according
to the standard wave equation, $\ddot v+k^2 v=0$, i.e.\ subhorizon GWs in the Jordan frame evolve
according to 
\[
h=h_{a}e^{i(kx-\omega t)}\,,\qquad h_{a}aM_{*}=\text{const},
\label{ampl}\]
where $h_{a}$ is the wave's amplitude. This result implies that $h_a$ is sensitive only to the ratio of the effective Planck mass and scale factors at emission and observation (see also the discussion of a phase shift in Ref.~\cite{Nishizawa:2017nef}).

In GR, the GW amplitude can be related to the luminosity distance $d_L$ of the source from the observer -- the potential evolution of $M_*$ is the only modification here, so that
\begin{equation}
h_a = \left(\frac{M_{*\text{,em}}}{M_{*\text{,obs}}}\right) \times h_s \,, \label{eq:five}
\end{equation}
where $h_s$  is the standard amplitude expression that, 
for merging binaries, can be approximated as
(see e.g. equation  (4.189) of \cite{maggiore2008gravitational})
\[
h_s=\frac{4}{d_L}\left(\frac{G\mathcal{M}_c}{c^2}\right)^{5/3}\left(\frac{\pi f_{\rm GW}}{c}\right)^{2/3}, \label{hgw2}
\]
with $\mathcal{M}_c$ the so-called chirp mass and $f_{\rm GW}$ the GW frequency measured by the observer.

The observable signal in the two polarizations $h_+,h_\times$ is finally obtained by multiplying $h$ by sinusoidal oscillations and by the factors $\cos i$ (for the $\times$ polarization) and the $(1+\cos^2 i)/2$ (for the $+$ polarization)  that depend on the inclination $i$ of the binary orbit with respect to the line of sight.

\begin{figure}
\includegraphics[width=0.5\textwidth]{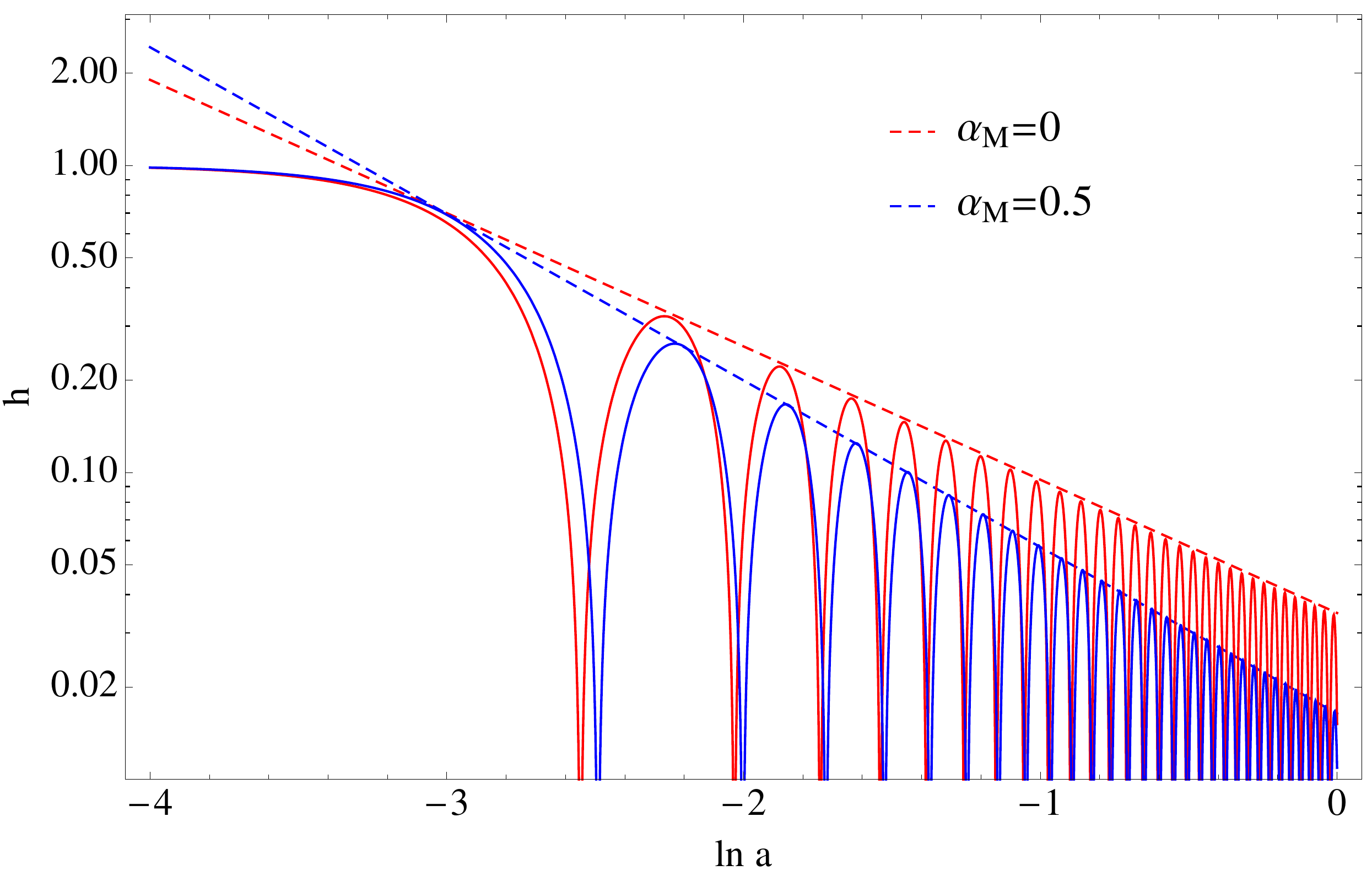}
\caption{Numerical evolution of the GW amplitude $|h|$ for illustrative choice of initial condition, parameter $\alpha_M=0,0.5$ and $k/H_0=100$ as a function of $\log_e a$. The dashed  lines represent the analytical result for the amplitude $a^{-1-\alpha_M/2}$. The phase shift is mainly driven by the different value of $\mu$ in Eq.~\eqref{eq:sho} and therefore a different horizon time, but is not relevant for GW wavelengths sourced by mergers from inspirals.}\label{fig1}
\end{figure}

As a concrete example, in the rest of this paper we assume for simplicity that $\alpha_M$ is constant in the region of observability (i.e.\ for $z\le 2$ roughly).\footnote{ The assumption $\alpha_M = \mathrm{const}$ is just for convenience; as Eq. (\ref{eq:five}) shows, the GW amplitude depends on the ratio of the Planck mass at emission and observation {\em independently} of the functional form of $\alpha_M$.} Then we have that,
\[
M_*\sim a^\frac{\alpha_M}{2},
\]
and
\begin{equation}
h_a =  (1+z)^{-\frac{\alpha_{M}}{2}}\times h_{s}. \label{hgw1}
\end{equation}
In Fig.~\ref{fig1}, we show  a  solution to \eqref{gweq} for a $\Lambda$CDM background and an illustrative choice of constant $\alpha_M$, comparing it to the evolution of the amplitude as given by \eqref{ampl}. Assuming the constancy of $\alpha_M$ all the way to recombination is excluded by e.g.\ the Planck analysis \cite{Ade:2015rim} at least when some extra assumptions about the model are taken, but it provides a clear illustration of the physics.

Equations (\ref{hgw1}) and (\ref{hgw2}) allow us to define an effective GW ``luminosity distance'' for the constant $\alpha_M$ case \cite{Nishizawa:2017nef}
\begin{equation}
d_{\rm GW}=(1+z)^{\frac{\alpha_{M}}{2}}d_{L}. \label{ddrel}
\end{equation}
Since the chirp mass, the inclination angle $i$, and the frequency $f_{\rm GW}$ can be measured independently of each other from the GW signal (see e.g. \cite{2005ApJ...629...15H}), GW experiments can measure directly the distance $d_{\rm GW}$.
In the next Section we discuss  possible future observations of $d_{\rm GW}$ and the constraints they can impose on $\alpha_M$.

An important issue concerns the actual GW waveform at the moment of emission. If gravity were modified, it is possible that the waveform would show qualitative differences compared to the corresponding one within GR as a result of radiation into a new degree of freedom (e.g.\ the fifth-force scalar), or a different deformation of the merging stars as a result of extra forces. Indeed, this would be yet another test of theories beyond GR in this context, however this does not come free of subtleties. Nonetheless, in gravity modifed at low curvatures, used as a model for late-time cosmology, the coupling of the fifth-force scalar is more suppressed the more compact the object, with black holes completely decoupled \cite{Hui:2012jb}. We thus expect no change in the waveform for black holes and only at most small corrections when dealing with neutron-star mergers. If gravity were \emph{in addition} modified at high curvatures, an accurate prediction of the waveform would require a good accounting of the various astrophysical factors such as the environment of the merger, the respective equations of state etc., which could potentially be degenerate with modified gravity effects (see e.g \cite{Berti:2015itd} for a detailed discussion). In this regard, modified gravity effects could also affect the predictions for quantities such as e.g.\ the chirp mass. 

On the other hand, the merging stars \emph{are} sensitive to the local value of $M_*$, which is a function of the scalar-field configuration at the site of the merger. This is not only determined by the cosmological value, but is affected by the profile of the mass in the galaxy and its surrounding environment. Screening mechanisms such as chameleon or Vainshtein, make this effect of mass distribution on the scalar large, rather than a small perturbation. When the emitting galaxies are screened, the local $M_*$ will still drift together with cosmology, although with a smaller amplitude and there will now be a scatter of values from different galaxies located at the same redshift. This would result in a scatter in the effective GW luminosity distance, reducing the sensitivity. This issue deserves a more thorough investigation, and the development of numerical simulations of mergers within modified gravity.


\section{Observing $\alpha_M$}

Given the result of the previous section, we now discuss how it can be applied to measure the time variation of the Planck mass through the parameter $\alpha_{M}$. Taking the $\log$ of Eq. (\ref{ddrel}) we have that,
\begin{equation}
\frac{2}{\log(1+z)}(\log d_{\rm GW}-\log d_{L})=\alpha_{M}.
\end{equation}
Then, assuming all variables to be Gaussian-distributed and statistically uncorrelated, the error on $\alpha_{M}$ can be estimated as
\begin{equation}
\sigma_{\alpha_{M}}=\frac{2}{\log(1+z)}\sqrt{\left(\frac{\sigma_{\rm GW}^{2}}{d_{\rm GW}^{2}}+\frac{\sigma_{L}^{2}}{d_{L}^{2}}\right)},\label{sam}
\end{equation}
where $\sigma^2_{\rm GW},\sigma^2_L$ are the variances of $d_{\rm GW}$ and $d_L$, respectively.
We neglect the error on $z$ because it is likely to be well below the errors on the other quantities. 

LIGO will be likely to obtain optical counterparts only at very low redshifts. For instance, the event reported in \cite{TheLIGOScientific:2017qsa} occurred at $z\approx 0.01$. It is then clear from Eq. (\ref{sam}) that the error on $\alpha_M$ is $\sim 200$ times the combined relative error of the distances and therefore very weak (see Ref. \cite{Arai:2017hxj}).%
\footnote{Although we note that a rapidly running effective Planck mass at very low redshifts would lead to a difference in the measurement of $H_0$ from GW and e.g.\ supernovae.} 
We therefore focus on future prospects with the LISA satellite.\footnote{www.elisascience.org}

LISA will measure $d_{\rm GW}$ up to very high redshifts. In particular, Ref. \cite{2016JCAP...04..002T} has shown that massive black-hole binaries (MBHB) can provide in 5 years' operation 30-50 identifiable optical counterparts distributed between $z=1$ and $z=8$, 5-15 of which within $z\le2$. From Ref. \cite{2016JCAP...04..002T}  (see e.g. their Fig. 1), an error of 5\% on $d_{\rm GW}  (z\approx 1.5)$ seems feasible. More stringent limits, down to  1\% or better, are quoted in \cite{2005ApJ...629...15H}. 

The luminosity distance $d_{L}$ might be measurable with supernovae Ia and baryon acoustic oscillations (BAO) up to say $z=2$. Although $d_L$ could be measured in the future to higher redshifts, for instance with real-time cosmology \cite{Quercellini:2010zr}, we restrict here ourselves to  measurements around $z\approx 1.5$. With $N$ supernovae Ia at this redshift, each with magnitude error $\Delta m$, one obtains a relative error on $d_L$ roughly equal to
\[
\frac{\Delta d_L}{d_L}\approx \frac{\log 10}{5}\frac{\Delta m}{\sqrt{N}}.
\]
Ref.\ \cite{2017arXiv171000844R} analyzed recently 15 supernovae Ia at $z>1$, nine of which within $1.5<z<2.3$, with $\Delta m\approx 0.2$. Taking indicatively 10 supernovae Ia at $z\approx 1.5$, we find a relative error of 3\% for $d_L(z\approx 1.5)$, which is indeed close to the values reported in \cite{2017arXiv171000844R}. BAO measurements with SKA and Euclid should reach 1\% accuracy at this redshift (see e.g. \cite{2015aska.confE..24B}, Fig. 3).

Taking now the conservative estimates of 5\% and 3\% for $d_{GW},d_L$, respectively, we obtain
\begin{equation}
\sigma_{\alpha_{M}}\approx 0.13.
\end{equation}
This improves to 0.03 with the optimistic estimates (1\% for both distances). 

These estimates can be significantly improved in two ways, at higher and at lower redshifts. At redshifts higher than 2, several GWs will be detected by  LISA, but at the moment we lack reliable distance indicators for $d_L$. At redshifts between 0.1 and 1, conversely, we have very good distance indicators, but a dearth of strong GW sources (in particular MBHB) detectable by LISA. The proposed Big Bang Observer (BBO) \cite{2009PhRvD..80j4009C} would be able to improve in this range, reaching a sensitivity on $d_{\rm GW}$ at the level of 0.1\%.
Below $z=0.1$, the $\log (1+z)$ denominator in Eq. (\ref{sam}) weakens the constraints below the threshold of interest.

In conclusion, we have discussed that GWs from optically identified sources can probe not only the propagation speed of GWs (through the effective parameter $\alpha_T$), but also a possible non-minimal, conformal coupling between the scalar sector and curvature through the time variation of the Planck mass, the latter being parametrised by the parameter $\alpha_M$. This result holds without any prior assumption about the particular cosmological model, and to a precision comparable or (with the BBO) superior to current tests.

\begin{acknowledgments}
\emph{Acknowledgements.}  
During the final stages of preparation of this manuscript, ref.~\cite{Belgacem:2017ihm} appeared with similar conclusions to this work.	
We thank Michele Maggiore and Filippo Vernizzi for useful discussions. The work of L.A.~is supported
by the DFG through TRR33 ``The  Dark Universe''. M.K.~acknowledges
funding by the Swiss National Science Foundation. I.S.~and I.D.S.~are supported by European Structural and Investment Funds and Czech Ministry of Education, Youth and Sports (Project CoGraDS -- CZ.02.1.01/0.0/0.0/15\_003/0000437).
\end{acknowledgments}

\bibliographystyle{utcaps}
\bibliography{AnisoRefs}

\providecommand{\href}[2]{#2}\begingroup\raggedright\begin{thebibliography}{10}

\bibitem{TheLIGOScientific:2017qsa}
{\bfseries Virgo, LIGO Scientific} Collaboration, B.~P. Abbott {\em et~al.},
  ``{GW170817: Observation of Gravitational Waves from a Binary Neutron Star
  Inspiral},'' \href{http://dx.doi.org/10.1103/PhysRevLett.119.161101}{{\em
  Phys. Rev. Lett.} {\bfseries 119} no.~16, (2017) 161101},
\href{http://arxiv.org/abs/1710.05832}{{\ttfamily arXiv:1710.05832 [gr-qc]}}.

\bibitem{2041-8205-848-2-L13}
B.~P. Abbott and et~al, ``{Gravitational Waves and Gamma-Rays from a Binary
  Neutron Star Merger: GW170817 and GRB 170817A},'' {\em The Astrophysical
  Journal Letters} {\bfseries 848} no.~2, (2017) L13.
  \url{http://stacks.iop.org/2041-8205/848/i=2/a=L13}.

\bibitem{PhysRevD.77.103513}
S.~F. Daniel, R.~R. Caldwell, A.~Cooray, and A.~Melchiorri, ``Large scale
  structure as a probe of gravitational slip,''
  \href{http://dx.doi.org/10.1103/PhysRevD.77.103513}{{\em Phys. Rev. D}
  {\bfseries 77} (May, 2008) 103513}.
  \url{https://link.aps.org/doi/10.1103/PhysRevD.77.103513}.

\bibitem{Saltas:2014dha}
I.~D. Saltas, I.~Sawicki, L.~Amendola, and M.~Kunz, ``{Anisotropic Stress as a
  Signature of Nonstandard Propagation of Gravitational Waves},''
  \href{http://dx.doi.org/10.1103/PhysRevLett.113.191101}{{\em Phys. Rev.
  Lett.} {\bfseries 113} no.~19, (2014) 191101},
\href{http://arxiv.org/abs/1406.7139}{{\ttfamily arXiv:1406.7139
  [astro-ph.CO]}}.

\bibitem{Sawicki:2016klv}
I.~Sawicki, I.~D. Saltas, M.~Motta, L.~Amendola, and M.~Kunz, ``{Nonstandard
  gravitational waves imply gravitational slip: On the difficulty of partially
  hiding new gravitational degrees of freedom},''
  \href{http://dx.doi.org/10.1103/PhysRevD.95.083520}{{\em Phys. Rev.}
  {\bfseries D95} no.~8, (2017) 083520},
\href{http://arxiv.org/abs/1612.02002}{{\ttfamily arXiv:1612.02002
  [astro-ph.CO]}}.

\bibitem{Amendola:2012ys}
{\bfseries Euclid Theory Working Group} Collaboration, L.~Amendola {\em
  et~al.}, ``{Cosmology and fundamental physics with the Euclid satellite},''
  {\em Living Rev.Rel.} {\bfseries 16} (2013) 6,
\href{http://arxiv.org/abs/1206.1225}{{\ttfamily arXiv:1206.1225
  [astro-ph.CO]}}.

\bibitem{Horndeski:1974wa}
G.~W. Horndeski, ``{Second-order scalar-tensor field equations in a
  four-dimensional space},''
\href{http://dx.doi.org/10.1007/BF01807638}{{\em Int.J.Theor.Phys.} {\bfseries
  10} (1974) 363--384}.

\bibitem{Deffayet:2011gz}
C.~Deffayet, X.~Gao, D.~A. Steer, and G.~Zahariade, ``{From k-essence to
  generalised Galileons},'' \href{http://arxiv.org/abs/1103.3260}{{\ttfamily
  arXiv:1103.3260 [hep-th]}}.

\bibitem{Kobayashi:2011nu}
T.~Kobayashi, M.~Yamaguchi, and J.~Yokoyama, ``{Generalized G-inflation:
  Inflation with the most general second-order field equations},''
  \href{http://dx.doi.org/10.1143/PTP.126.511}{{\em Prog. Theor. Phys.}
  {\bfseries 126} (2011) 511--529},
\href{http://arxiv.org/abs/1105.5723}{{\ttfamily arXiv:1105.5723 [hep-th]}}.

\bibitem{DeFelice:2011bh}
A.~De~Felice and S.~Tsujikawa, ``{Conditions for the cosmological viability of
  the most general scalar-tensor theories and their applications to extended
  Galileon dark energy models},''
  \href{http://dx.doi.org/10.1088/1475-7516/2012/02/007}{{\em JCAP} {\bfseries
  1202} (2012) 007},
\href{http://arxiv.org/abs/1110.3878}{{\ttfamily arXiv:1110.3878 [gr-qc]}}.

\bibitem{Jacobson:2000xp}
T.~Jacobson and D.~Mattingly, ``{Gravity with a dynamical preferred frame},''
  \href{http://dx.doi.org/10.1103/PhysRevD.64.024028}{{\em Phys.Rev.}
  {\bfseries D64} (2001) 024028},
\href{http://arxiv.org/abs/gr-qc/0007031}{{\ttfamily arXiv:gr-qc/0007031
  [gr-qc]}}.

\bibitem{Heisenberg:2014kea}
L.~Heisenberg, R.~Kimura, and K.~Yamamoto, ``{Cosmology of the proxy theory to
  massive gravity},''
\href{http://arxiv.org/abs/1403.2049}{{\ttfamily arXiv:1403.2049 [hep-th]}}.

\bibitem{Tasinato:2014eka}
G.~Tasinato, ``{Cosmic Acceleration from Abelian Symmetry Breaking},''
  \href{http://dx.doi.org/10.1007/JHEP04(2014)067}{{\em JHEP} {\bfseries 04}
  (2014) 067},
\href{http://arxiv.org/abs/1402.6450}{{\ttfamily arXiv:1402.6450 [hep-th]}}.

\bibitem{Gubitosi:2012hu}
G.~Gubitosi, F.~Piazza, and F.~Vernizzi, ``{The Effective Field Theory of Dark
  Energy},'' \href{http://dx.doi.org/10.1088/1475-7516/2013/02/032}{{\em JCAP}
  {\bfseries 1302} (2013) 032},
\href{http://arxiv.org/abs/1210.0201}{{\ttfamily arXiv:1210.0201 [hep-th]}}.

\bibitem{Gleyzes:2013ooa}
J.~Gleyzes, D.~Langlois, F.~Piazza, and F.~Vernizzi, ``{Essential Building
  Blocks of Dark Energy},''
  \href{http://dx.doi.org/10.1088/1475-7516/2013/08/025}{{\em JCAP} {\bfseries
  1308} (2013) 025},
\href{http://arxiv.org/abs/1304.4840}{{\ttfamily arXiv:1304.4840 [hep-th]}}.

\bibitem{Bloomfield:2013efa}
J.~Bloomfield, ``{A Simplified Approach to General Scalar-Tensor Theories},''
\href{http://arxiv.org/abs/1304.6712}{{\ttfamily arXiv:1304.6712
  [astro-ph.CO]}}.

\bibitem{Bellini:2014fua}
E.~Bellini and I.~Sawicki, ``{Maximal freedom at minimum cost: linear
  large-scale structure in general modifications of gravity},''
  \href{http://dx.doi.org/10.1088/1475-7516/2014/07/050}{{\em JCAP} {\bfseries
  1407} (2014) 050},
\href{http://arxiv.org/abs/1404.3713}{{\ttfamily arXiv:1404.3713
  [astro-ph.CO]}}.

\bibitem{Zumalacarregui:2013pma}
M.~Zumalac\'{a}rregui and J.~Garc\'{i}a-Bellido, ``{Transforming gravity: from
  derivative couplings to matter to second-order scalar-tensor theories beyond
  the Horndeski Lagrangian},''
\href{http://arxiv.org/abs/1308.4685}{{\ttfamily arXiv:1308.4685 [gr-qc]}}.

\bibitem{Gleyzes:2014dya}
J.~Gleyzes, D.~Langlois, F.~Piazza, and F.~Vernizzi, ``{Healthy theories beyond
  Horndeski},''
\href{http://arxiv.org/abs/1404.6495}{{\ttfamily arXiv:1404.6495 [hep-th]}}.

\bibitem{Lagos:2017hdr}
M.~Lagos, E.~Bellini, J.~Noller, P.~G. Ferreira, and T.~Baker, ``{A general
  theory of linear cosmological perturbations: stability conditions, the
  quasistatic limit and dynamics},''
\href{http://arxiv.org/abs/1711.09893}{{\ttfamily arXiv:1711.09893 [gr-qc]}}.

\bibitem{Creminelli:2017sry}
P.~Creminelli and F.~Vernizzi, ``{Dark Energy after GW170817 and GRB170817A},''
  \href{http://dx.doi.org/10.1103/PhysRevLett.119.251302}{{\em Phys. Rev.
  Lett.} {\bfseries 119} no.~25, (2017) 251302},
\href{http://arxiv.org/abs/1710.05877}{{\ttfamily arXiv:1710.05877
  [astro-ph.CO]}}.

\bibitem{Ezquiaga:2017ekz}
J.~M. Ezquiaga and M.~Zumalac\'{a}rregui, ``{Dark Energy After GW170817: Dead
  Ends and the Road Ahead},''
  \href{http://dx.doi.org/10.1103/PhysRevLett.119.251304}{{\em Phys. Rev.
  Lett.} {\bfseries 119} no.~25, (2017) 251304},
\href{http://arxiv.org/abs/1710.05901}{{\ttfamily arXiv:1710.05901
  [astro-ph.CO]}}.

\bibitem{Sakstein:2017xjx}
J.~Sakstein and B.~Jain, ``{Implications of the Neutron Star Merger GW170817
  for Cosmological Scalar-Tensor Theories},''
  \href{http://dx.doi.org/10.1103/PhysRevLett.119.251303}{{\em Phys. Rev.
  Lett.} {\bfseries 119} no.~25, (2017) 251303},
\href{http://arxiv.org/abs/1710.05893}{{\ttfamily arXiv:1710.05893
  [astro-ph.CO]}}.

\bibitem{Baker:2017hug}
T.~Baker, E.~Bellini, P.~G. Ferreira, M.~Lagos, J.~Noller, and I.~Sawicki,
  ``{Strong constraints on cosmological gravity from GW170817 and GRB
  170817A},'' \href{http://dx.doi.org/10.1103/PhysRevLett.119.251301}{{\em
  Phys. Rev. Lett.} {\bfseries 119} no.~25, (2017) 251301},
\href{http://arxiv.org/abs/1710.06394}{{\ttfamily arXiv:1710.06394
  [astro-ph.CO]}}.

\bibitem{Bettoni:2016mij}
D.~Bettoni, J.~M. Ezquiaga, K.~Hinterbichler, and M.~Zumalac\'{a}rregui,
  ``{Speed of Gravitational Waves and the Fate of Scalar-Tensor Gravity},''
  \href{http://dx.doi.org/10.1103/PhysRevD.95.084029}{{\em Phys. Rev.}
  {\bfseries D95} no.~8, (2017) 084029},
\href{http://arxiv.org/abs/1608.01982}{{\ttfamily arXiv:1608.01982 [gr-qc]}}.

\bibitem{Lombriser:2016yzn}
L.~Lombriser and N.~A. Lima, ``{Challenges to Self-Acceleration in Modified
  Gravity from Gravitational Waves and Large-Scale Structure},''
  \href{http://dx.doi.org/10.1016/j.physletb.2016.12.048}{{\em Phys. Lett.}
  {\bfseries B765} (2017) 382--385},
\href{http://arxiv.org/abs/1602.07670}{{\ttfamily arXiv:1602.07670
  [astro-ph.CO]}}.

\bibitem{Amendola:2017orw}
L.~Amendola, M.~Kunz, I.~D. Saltas, and I.~Sawicki, ``{The fate of large-scale
  structure in modified gravity after GW170817 and GRB170817A},''
\href{http://arxiv.org/abs/1711.04825}{{\ttfamily arXiv:1711.04825
  [astro-ph.CO]}}.

\bibitem{Arai:2017hxj}
S.~Arai and A.~Nishizawa, ``{Generalized framework for testing gravity with
  gravitational-wave propagation. II. Constraints on Horndeski theory},''
\href{http://arxiv.org/abs/1711.03776}{{\ttfamily arXiv:1711.03776 [gr-qc]}}.

\bibitem{2017arXiv170701866B}
A.~{Bonanno} and H.-E. {Fr{\"o}hlich}, ``{A new helioseismic constraint on a
  cosmic-time variation of G},'' {\em ArXiv e-prints} (July, 2017) ,
  \href{http://arxiv.org/abs/1707.01866}{{\ttfamily arXiv:1707.01866
  [astro-ph.SR]}}.

\bibitem{2004PhRvL..92q1301C}
C.~J. {Copi}, A.~N. {Davis}, and L.~M. {Krauss}, ``{New Nucleosynthesis
  Constraint on the Variation of G},''
  \href{http://dx.doi.org/10.1103/PhysRevLett.92.171301}{{\em Physical Review
  Letters} {\bfseries 92} no.~17, (Apr., 2004) 171301},
  \href{http://arxiv.org/abs/astro-ph/0311334}{{\ttfamily astro-ph/0311334}}.

\bibitem{Pettorino:2014bka}
V.~Pettorino and L.~Amendola, ``{Friction in Gravitational Waves: a test for
  early-time modified gravity},''
  \href{http://dx.doi.org/10.1016/j.physletb.2015.02.007}{{\em Phys. Lett.}
  {\bfseries B742} (2015) 353--357},
\href{http://arxiv.org/abs/1408.2224}{{\ttfamily arXiv:1408.2224
  [astro-ph.CO]}}.

\bibitem{Ade:2015rim}
{\bfseries Planck} Collaboration, P.~A.~R. Ade {\em et~al.}, ``{Planck 2015
  results. XIV. Dark energy and modified gravity},''
  \href{http://dx.doi.org/10.1051/0004-6361/201525814}{{\em Astron. Astrophys.}
  {\bfseries 594} (2016) A14},
\href{http://arxiv.org/abs/1502.01590}{{\ttfamily arXiv:1502.01590
  [astro-ph.CO]}}.

\bibitem{Nishizawa:2017nef}
A.~Nishizawa, ``{Generalized framework for testing gravity with
  gravitational-wave propagation. I. Formulation},''
\href{http://arxiv.org/abs/1710.04825}{{\ttfamily arXiv:1710.04825 [gr-qc]}}.

\bibitem{maggiore2008gravitational}
M.~Maggiore, {\em Gravitational Waves: Volume 1: Theory and Experiments}.
\newblock Gravitational Waves. OUP Oxford, 2008.
\newblock \url{https://books.google.de/books?id=AqVpQgAACAAJ}.

\bibitem{2005ApJ...629...15H}
D.~E. {Holz} and S.~A. {Hughes}, ``{Using Gravitational-Wave Standard
  Sirens},'' \href{http://dx.doi.org/10.1086/431341}{{\em Ap. J.} {\bfseries
  629} (Aug., 2005) 15--22},
  \href{http://arxiv.org/abs/astro-ph/0504616}{{\ttfamily astro-ph/0504616}}.

\bibitem{Hui:2012jb}
L.~Hui and A.~Nicolis, ``{Proposal for an Observational Test of the Vainshtein
  Mechanism},'' \href{http://dx.doi.org/10.1103/PhysRevLett.109.051304}{{\em
  Phys. Rev. Lett.} {\bfseries 109} (2012) 051304},
\href{http://arxiv.org/abs/1201.1508}{{\ttfamily arXiv:1201.1508
  [astro-ph.CO]}}.

\bibitem{Berti:2015itd}
E.~Berti {\em et~al.}, ``{Testing General Relativity with Present and Future
  Astrophysical Observations},''
  \href{http://dx.doi.org/10.1088/0264-9381/32/24/243001}{{\em Class. Quant.
  Grav.} {\bfseries 32} (2015) 243001},
\href{http://arxiv.org/abs/1501.07274}{{\ttfamily arXiv:1501.07274 [gr-qc]}}.

\bibitem{2016JCAP...04..002T}
N.~{Tamanini}, C.~{Caprini}, E.~{Barausse}, A.~{Sesana}, A.~{Klein}, and
  A.~{Petiteau}, ``{Science with the space-based interferometer eLISA. III:
  probing the expansion of the universe using gravitational wave standard
  sirens},'' \href{http://dx.doi.org/10.1088/1475-7516/2016/04/002}{{\em JCAP}
  {\bfseries 4} (Apr., 2016) 002},
  \href{http://arxiv.org/abs/1601.07112}{{\ttfamily arXiv:1601.07112}}.

\bibitem{Quercellini:2010zr}
C.~Quercellini, L.~Amendola, A.~Balbi, P.~Cabella, and M.~Quartin, ``{Real-time
  Cosmology},'' \href{http://dx.doi.org/10.1016/j.physrep.2012.09.002}{{\em
  Phys. Rept.} {\bfseries 521} (2012) 95--134},
\href{http://arxiv.org/abs/1011.2646}{{\ttfamily arXiv:1011.2646
  [astro-ph.CO]}}.

\bibitem{2017arXiv171000844R}
A.~G. {Riess}, S.~A. {Rodney}, D.~M. {Scolnic}, D.~L. {Shafer}, L.-G.
  {Strolger}, H.~C. {Ferguson}, M.~{Postman}, O.~{Graur}, D.~{Maoz}, S.~W.
  {Jha}, B.~{Mobasher}, S.~{Casertano}, B.~{Hayden}, A.~{Molino}, J.~{Hjorth},
  P.~M. {Garnavich}, D.~O. {Jones}, R.~P. {Kirshner}, A.~M. {Koekemoer}, N.~A.
  {Grogin}, G.~{Brammer}, S.~{Hemmati}, M.~{Dickinson}, P.~M. {Challis},
  S.~{Wolff}, K.~I. {Clubb}, A.~V. {Filippenko}, H.~{Nayyeri}, U.~{Vivian},
  D.~C. {Koo}, S.~M. {Faber}, D.~{Kocevski}, L.~{Bradley}, and D.~{Coe},
  ``{Type Ia Supernova Distances at z > 1.5 from the Hubble Space Telescope
  Multi-Cycle Treasury Programs: The Early Expansion Rate},'' {\em ArXiv
  e-prints} (Oct., 2017) , \href{http://arxiv.org/abs/1710.00844}{{\ttfamily
  arXiv:1710.00844}}.

\bibitem{2015aska.confE..24B}
P.~{Bull}, S.~{Camera}, A.~{Raccanelli}, C.~{Blake}, P.~{Ferreira},
  M.~{Santos}, and D.~J. {Schwarz}, ``{Measuring baryon acoustic oscillations
  with future SKA surveys},'' {\em Advancing Astrophysics with the Square
  Kilometre Array (AASKA14)} (Apr., 2015) 24,
  \href{http://arxiv.org/abs/1501.04088}{{\ttfamily arXiv:1501.04088}}.

\bibitem{2009PhRvD..80j4009C}
C.~{Cutler} and D.~E. {Holz}, ``{Ultrahigh precision cosmology from
  gravitational waves},''
  \href{http://dx.doi.org/10.1103/PhysRevD.80.104009}{{\em Phys. Rev. D}
  {\bfseries 80} no.~10, (Nov., 2009) 104009},
  \href{http://arxiv.org/abs/0906.3752}{{\ttfamily arXiv:0906.3752
  [astro-ph.CO]}}.

\bibitem{Belgacem:2017ihm}
E.~Belgacem, Y.~Dirian, S.~Foffa, and M.~Maggiore, ``{The gravitational-wave
  luminosity distance in modified gravity theories},''
\href{http://arxiv.org/abs/1712.08108}{{\ttfamily arXiv:1712.08108
  [astro-ph.CO]}}.

\end{thebibliography}\endgroup

\end{document}